\documentclass[twocolumn,showpacs,preprintnumbers,amsmath,amssymb,floatfix]{revtex4}

\usepackage{graphicx}
\usepackage{dcolumn}
\usepackage{bm}
\usepackage{subfigure} 

\begin{document}

\preprint{}

\title{Direct numerical simulation of homogeneous nucleation and growth in a phase-field model using cell dynamics method}

\author{Masao Iwamatsu}
\email{iwamatsu@ph.ns.musashi-tech.ac.jp}
\affiliation{
Department of Physics, General Education Center,
Musashi Institute of Technology,
Setagaya-ku, Tokyo 158-8557, Japan
}%


\date{\today}

\begin{abstract}
Homogeneous nucleation and growth in a simplest two-dimensional phase field model is numerically studied using the cell dynamics method.  Whole process from nucleation to growth is simulated and is shown to follow closely the Kolmogorov-Johnson-Mehl-Avrami (KJMA) scenario of phase transformation.  Specifically the time evolution of the volume fraction of new stable phase is found to follow closely the KJMA formula.  By fitting the KJMA formula directly to the simulation data, not only the Avrami exponent but the magnitude of nucleation rate and, in particular, of incubation time are quantitatively studied.  The modified Avrami plot is also used to verify the derived KJMA parameters.  It is found that the Avrami exponent is close to the ideal theoretical value $m=3$.  The temperature dependence of nucleation rate follows the activation-type behavior expected from the classical nucleation theory.  On the other hand, the temperature dependence of incubation time does not follow the exponential activation-type behavior.  Rather the incubation time is inversely proportional to the temperature predicted from the theory of Shneidman and Weinberg [J. Non-Cryst. Solids {\bf 160}, 89 (1993)].  A need to restrict thermal noise in simulation to deduce correct Avrami exponent is also discussed.    
\end{abstract}

\maketitle

\section{Introduction}
\label{sec:level1}

The dynamics of phase transformation by nucleation and growth of a new stable phase from a metastable phase is a very old problem, which has been studied for more than half a century~\cite{Christian,Kelton,Oxtoby} from a fundamental point of view as well as from technological interests.  Experimentally~\cite{Price,Weinberg}, the dynamics of phase transformation has been believed to follow the classical Kolmogorov-Johnson-Mehl-Avrami (KJMA) picture of nucleation and growth~\cite{Christian,Kolmogorov,Johnson,Avrami}.  According to their picture, the phase transformation proceeds via the nucleation and subsequent growth of nuclei by the interface-limited growth. Therefore, the classical nucleation theory and the steady-state growth are assumed.  Their picture is integrated into the well-known KJMA formula for the fraction of transformed volume.

Theoretically, on the other hand, the various variant of phase-field~\cite{Jou,Castro,Granasy} model has been routinely used to study the dynamics of phase transformation.  Essentially similar models called Cahn-Hilliard model~\cite{Cahn} and time-dependent Ginzburg-Landau model~\cite{Langer,Valls} have also been used extensively.  It is not obvious, however, that the dynamics of phase-field model is in accord with the KJMA picture. In particular, the time scale of phase transformation including the transient nucleation has not been fully studied. For more than a decade ago, Valls and Mazenko~\cite{Valls} studied the dynamics of phase transformation in a phase-field model.  They studied nucleation and growth rather phenomenological way and did not pay attention to the connection to KJMA formula.  Very recently, Jou and Lusk~\cite{Jou}, and Castro~\cite{Castro} studied the connection between phase-field model and KJMA formula.  However, the former introduced critical nucleus (seed of new phase) artificially and the latter paid more attention to the heterogeneous nucleation. Therefore, the nucleation process is artificially controlled in their studies~\cite{Jou,Castro}.  Very recently, Gr\'an\'asy {\it et al.}~\cite{Granasy2} extensively studied the validity of the KJMA picture in the phase-field model of spherulite.  They paid more attention to the Avrami exponent rather than the time scale of phase transition. Therefore, virtually there has been no detailed study of the time scale of phase transition including both the nucleation and the growth in phase-field model on the same footing. 

In this paper, we will use the phase-field model with thermal noise to study the homogeneous nucleation and the growth in a unified manner.  In Sec. II we present a short review of the classical KJMA picture of nucleation and growth.  In Sec. III, we present the phase-field model and the cell dynamics method.  Section IV is devoted to the results of numerical simulations. Finally Sec. V is devoted to the conclusion.

\section{Classical KJMA picture of nucleation and growth}
The time evolution of the volume fraction $f$ of transformed volume for two-dimensional system is predicted from KJMA (Kolmogorov-Johnson-Mehl-Avrami)~\cite{Christian,Kolmogorov,Johnson,Avrami,Shneidman1} theory as
\begin{equation}
f = 1 - \exp\left[-\pi\Omega_{2}(t)\right],
\label{eq:3.1}
\end{equation}
with
\begin{equation}
\Omega_{2}(t) =\int_{0}^{t}I(t')R^{2}(t;t')dt',
\label{eq:3.2}
\end{equation}
where $I(t)$ is the (time-dependent) nucleation rate and $R(t;t')$ is the radius of nucleus at time $t$ that was nucleated at $t'$. 

These equations are further simplified if~\cite{Shneidman1}
\begin{enumerate}
\renewcommand{\labelenumi}{(\roman{enumi})}
\item the interfacial velocity $v_{s}$ is time- and size-independent, and
\item the radius of critical nucleus $R_{*}$ is infinitesimally small, and
\item the nucleation rate (for critical nucleus) $I(t)$ is approximated by the (time-independent) steady state nucleation rate $I_{s}$,
\end{enumerate}
then we have a linear time dependence of radius of nuclei $R(t;t')=v_{s}(t-t')$, and the integral in Eq.~(\ref{eq:3.2}) can be calculated analytically to give the classical KJMA formula:
\begin{equation}
f = 1 - \exp\left(-\frac{\pi}{3}I_{s}v_{s}^{2}t^{3}\right).
\label{eq:3.3}
\end{equation}

However, if we take into account the fact that the radius of critical nucleus $R_{*}$ is finite, we have to shift the origin of time scale toward the past by the amount $t_{0}\simeq R_{*}/v_{s}$.  Then, we have to replace time $t$ by~\cite{Shneidman3}
\begin{equation}
t \Rightarrow t+t_{0}.
\label{eq:3.4}
\end{equation}
According to the theory of Shneidman and Weinberg~\cite{Shneidman1}, on the other hand, if we take into account the transient nucleation rate $I(t)$ as well as the size dependence of interfacial velocity $v$, we have to replace the origin of time $t$ toward future by
\begin{equation}
t \Rightarrow t\left[1- \frac{t_{0}}{t}\ln\left(\frac{t}{t_{0}}\frac{W_{*}}{k_{\rm B}T},\right)\right]
\label{eq:3.5}
\end{equation}
where $W_{*}$ is the energy barrier to form the critical nucleus and $T$ is the absolute temperature, since we have to wait for the nucleation rate as well as the interfacial velocity to reach the steady state values.  Therefore two contributions Eq.~(\ref{eq:3.4}) and Eq.~(\ref{eq:3.5}) have opposite signs and we have to replace time $t$ in Eq.~(\ref{eq:3.3}) by
\begin{equation}
t \Rightarrow t - t_{\rm inc},
\label{eq:3.6}
\end{equation}
where $t_{\rm inc}$ is usually called "incubation time", which now consists of two contributions Eq.~(\ref{eq:3.4}) and Eq.~(\ref{eq:3.5}):
\begin{equation}
t_{\rm inc}\simeq t_{0}\ln\left(\frac{t}{t_{0}}\frac{W_{*}}{k_{\rm B}T}\right)-t_{0}.
\label{eq:3.7}
\end{equation}
where the time $t\gg t_{0}$ is the timescale of the total process of phase transformation.  At low temperature (small $T$) or low undercooling (small $\epsilon$ and, therefore, large $W_{*}$), critical nucleus is rarely formed and the time necessary to reach the steady state would be long.  Then the first term of Eq.~(\ref{eq:3.7}) dominates and we would have a positive $t_{\rm inc}>0$.  Usual condition of the classical nucleation theory $W_{*}/k_{\rm B}T\simeq 40-70$~\cite{Kelton,Oxtoby} satisfies this condition. 

Furthermore, if we have anomalous power-law dependence of nucleation rate $I(t)\simeq I_{0}t^{\alpha}$ and interfacial velocity $v(t)\simeq v_{0}t^{\beta}$, then we have a generalized KJMA formula
\begin{equation}
f = 1 - \exp\left[-\left(\frac{t-t_{\rm inc}}{t_{\rm gr}}\right)^{m}\right],
\label{eq:3.9}
\end{equation}
where
\begin{equation}
m = 3+\alpha+2\beta
\label{eq:3.9a}
\end{equation}
is the so-called Avrami exponent, $t_{\rm inc}$ the incubation time, and $t_{\rm gr}$ is the growth time, which is determined from the nucleation rate $I_{0}$ and the interfacial velocity $v_{0}$ through
\begin{equation}
\frac{1}{t_{\rm gr}}=\left(\frac{\pi}{m}I_{0}v_{0}^{2}\right)^{1/m}.
\label{eq:3.10}
\end{equation}
The ideal case Eq.~(\ref{eq:3.3}) is given by the formula with $\alpha=\beta=0$ and $m=3$.

In the classical nucleation theory (CNT)~\cite{Christian,Kelton,Oxtoby}, the steady-state nucleation rate $I_{s}\simeq I_{0}$, that is the number of critical nuclei which appear per unit time and unit volume, is usually given by the activation form
\begin{equation}
I_{s}\propto \exp\left(-\frac{W_{*}}{k_{\rm B}T}\right),
\label{eq:3.11}
\end{equation}
where $W_{*}$ is the nucleation barrier of critical nucleus that appeared in Eq.~(\ref{eq:3.5}). Then, the growth time is given by the activation form:
\begin{equation}
t_{\rm gr} \propto \exp\left(\frac{W_{*}}{m k_{\rm B}T}\right),
\label{eq:3.11xx}
\end{equation}
Similarly, by replacing the time scale $t$ in Eq.~(\ref{eq:3.7}) by $t_{\rm gr}$ in Eq.~(\ref{eq:3.11xx}), we have 
\begin{equation}
t_{\rm inc}\simeq t_{0}\ln\left(\frac{t_{\rm gr}}{t_{0}}\right)
\propto \frac{W_{*}}{m k_{\rm B}T}.
\label{eq:3.11x}
\end{equation}
The two time-scales of phase transformation, the incubation time $t_{\rm inc}$ and the growth time $t_{\rm gr}$ will increase as the temperature $T$ is lowered.  But the temperature dependence is different.  The growth time increases exponentially and it increases much faster than the incubation time as the temperature is lowered.

The classical expression of the energy barrier $W_{*}$ for the two-dimensional circular nucleus is given by
\begin{equation}
W_{*}=\frac{\pi\sigma^{2}}{\Delta h}
\label{eq:3.12}
\end{equation}
with radius of critical nucleus $R_{*}$ given by
\begin{equation}
R_{*}=\frac{\sigma}{\Delta h}
\label{eq:3.12x}
\end{equation}
where $\sigma$ is the interfacial energy of nucleus and $\Delta h$ is the free energy difference between the metastable and the stable phase.  

In section IV we will check if this classical KJMA picture of nucleation and growth~\cite{Christian,Kolmogorov,Johnson,Avrami,Shneidman1}, in particular Eqs.~(\ref{eq:3.9}), (\ref{eq:3.11xx}) and (\ref{eq:3.11x}) are valid in the phase-field model using cell dynamics method. A similar study to test the validity of KJMA picture in Ising-type spin model was conducted by Shneideman {\it et al.}~\cite{Shneidman3} and Ramos {\it et al.}~\cite{Ramos}.

\section{Phase-Field Model and Cell Dynamics Method}
\label{sec:sec2}

In the phase-field model, the dynamics of phase transformation is described by the standard isothermal phase-field equation~\cite{Valls,Jou,Castro,Granasy} for continuous variables
\begin{equation}
\frac{\partial \psi}{\partial t}=-\frac{\delta  F}{\delta \psi},
\label{eq:2-1}
\end{equation}
where $\delta$ denotes the functional differentiation, $\psi$ is the {\it non-conserved} order parameter, and $F$ is the free energy functional. This free energy is usually written as the square-gradient form:
\begin{equation}
 F[\psi]=\frac{1}{2}\int \left[D(\nabla \psi)^{2}+h(\psi)\right]{\rm d}{\bf r}. 
\label{eq:2-2}
\end{equation}
The local part $h(\psi)$ of the free energy $F$ determines the bulk phase diagram and the value of the order parameter in equilibrium phases.  

In the cell dynamics method, the partial differential equation (\ref{eq:2-1}) is replaced by a finite difference equation in space and time in the form
\begin{equation}
\psi(t+1,n)=M[\psi(t,n)],
\label{eq:2-3}
\end{equation}
where the time $t$ is discrete and an integer, and the space is also discrete and is expressed by the integral site index $n$.  The mapping $M$ is given by
\begin{equation}
M[\psi(t,n)]=-f(\psi(t,n))+D\left[\ll\psi(t,n)\gg-\psi(t,n)\right],
\label{eq:2-4}
\end{equation}
where $f(\psi)=dh(\psi)/d\psi$, and the definition of $\ll\cdots\gg$ for the two-dimensional square grid is given in \cite{Oono,Puri}.  We use the map function $f(\psi)$ directly obtained from the free energy $h(\psi)$~\cite{Iwamatsu1,Iwamatsu2} instead of the standard $\tanh$ form originally used by Oono and Puri~\cite{Oono,Puri}, which is essential for studying the subtle nature of nucleation and growth when one phase is metastable and another is stable.

\begin{figure}[htbp]
\begin{center}
\includegraphics[width=0.90\linewidth]{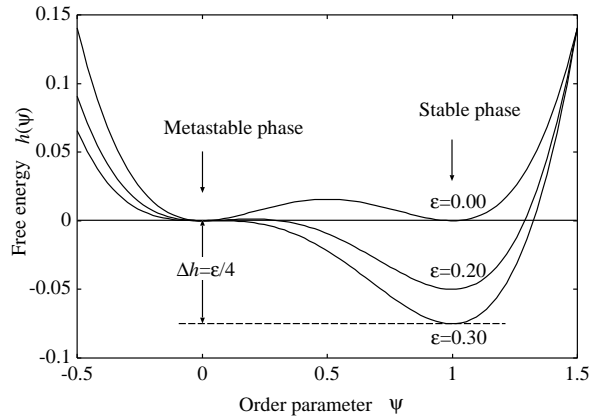}
\end{center}
\caption{
Model double-well free energy $h(\psi)$ defined by Eq.~(\ref{eq:2-8}).  The undercooling parameter $\epsilon$ determines the free energy difference $\Delta h=\epsilon/4$ between the depth of two wells.  The phase with $\psi=0$ is metastable while the one with $\psi=1$ is stable.
}
\label{fig:1}
\end{figure}

The local part of the free energy $h(\psi)$ we use is~\cite{Jou,Iwamatsu1}
\begin{equation}
h_(\psi) = \frac{1}{4}\psi^{2}(1-\psi)^{2} + \frac{3}{2}\epsilon\left(\frac{\psi^{3}}{3}-\frac{\psi^{2}}{2}\right).
\label{eq:2-8}
\end{equation}
This free energy is shown in Fig.~\ref{fig:1}, where one phase at $\psi=0$ is metastable while another phase at $\psi=1$ is stable. The free energy difference $\Delta h$ between the stable phase at $\psi=1$ and the metastable phase at $\psi=0$ is determined from the parameter $\epsilon$:
\begin{equation}
\Delta h = h(\psi=0)-h(\psi=1)=\frac{\epsilon}{4}.
\label{eq:2-9}
\end{equation}
We will use the terminology undercooling to represent $\epsilon$. The metastable phase at $\psi=0$ becomes unstable when $\epsilon =1/3$, which defines the spinodal.  

The interfacial energy $\sigma$ in Eqs.~(\ref{eq:3.12}) and (\ref{eq:3.12x}) can be estimated from the interfacial profile calculated from the Ginzburg-Landau equation $\delta F/\delta \psi=0$ for the order parameter $\psi$ at the two-phase coexistence ($\epsilon=0$)~\cite{Iwamatsu3}.  Then the $\epsilon$-dependence of the surface tension $\sigma$ could be neglected, and the energy barrier $W_{*}$ and the radius of critical nucleus in Eqs.~(\ref{eq:3.12}) and (\ref{eq:3.12x}) becomes inversely proportional to the under cooling $\epsilon$:
\begin{equation}
W_{*} \propto \frac{1}{\epsilon},\;\;R_{*} \propto \frac{1}{\epsilon}
\label{eq:3.12y}
\end{equation}
These formulae are incorrect when the spinodal is approached when $\epsilon\rightarrow 1/3$ as the surface tension will vanishes ($\sigma \rightarrow 0$) at the spinodal.  Then, both the energy barrier $W_{*}$ and the radius $R_{*}$ of critical nucleus vanish at the spinodal.  Equation (\ref{eq:3.12y}) cannot be used near the spinodal.

In the previous paper~\cite{Iwamatsu1}, using this phase-field model and the cell dynamics method, we could successfully simulate the interface-limited growth of a single nucleus~\cite{Chan} with a constant growth velocity $v$ which is close to the theoretical prediction~\cite{Chan,Iwamatsu1}
\begin{equation}
v=\frac{1}{2}\sqrt{\frac{D}{2}}3\epsilon.
\label{eq:3.13}
\end{equation} 
The interfacial velocity $v$ is constant during evolution and does not depend on the temperature, that is in accord with the assumption (i) of the original KJMA formula Eq.~(\ref{eq:3.3}).  Therefore, the temperature dependence of growth time $t_{\rm gr}$ in Eq.~(\ref{eq:3.10}) comes solely from the temperature dependence of nucleation rate $I_{s}$ given by Eq.~(\ref{eq:3.11}).  

By introducing the critical nucleus artificially~\cite{Iwamatsu1,Jou}, the growth of the multiple of nuclei with the site-saturation condition as well as with the continuous homogeneous nucleation condition could also be studied. In this case, the critical nucleus is not introduced in the region where the phase transformation has  already started.  Introduction of an extra nucleus on the boundary of phase transformed region will accelerate the phase transformation and will result in the increase of apparent nucleation rate. With such a care, the phase-field model can reproduce the theoretically predicted ideal Avrami exponent $m\simeq 2$ for the site saturation case and $m\simeq 3$ for the continuous nucleation case~\cite{Iwamatsu1}.  

In order to simulate not only the growth process but the nucleation process in the phase field model, we add thermal noise in this paper in Eq.~(\ref{eq:2-3}):
\begin{equation}
\psi(t+1,n)=M[\psi(t,n)] + \xi(t,n)
\label{eq:2-3x}
\end{equation}
The thermal noise $\xi(t)$ should be added to only those regions where the phase transformation has not yet started.  This process is simulated by adding thermal noise $\xi(t,n)$ only when the phase field is smaller than some cutoff value $\psi_{c}$.  Therefore stochastic Eq.~(\ref{eq:2-3x}) is used only when $\psi(t,n)<\psi_{c}$ otherwise deterministic Eq.~(\ref{eq:2-3}) is used.  We set $\psi_{c}=0.2$ thought this paper.  Too small $\psi_{c}<0.2$ suppress nucleation completely, while larger $\psi_{c}>0.2$ will enhance the nucleation rate and will lead to larger Avrami exponent $m>3$ and non-linear behavior of nucleation and growth parameters.  The limitation of this ad hock method will be discussed later in section \ref{sec:4}.

The thermal noise $\xi(t,n)$ is related to the absolute temperature $T$ from the fluctuation-dissipation theorem as
\begin{equation}
\left<\xi(t,n)\xi(t',n')\right>=k_{\rm B}T\delta_{n,n'}\delta_{t,t'}.
\label{eq:2-7}
\end{equation}
In this paper, we will use a uniform random number ranging from $-\sqrt{\tau}$ to $+\sqrt{\tau}$ for the thermal noise $\xi(t,n)$~\cite{Oono,Puri}.  Then, the parameter $\tau$ is proportional to the absolute temperature:
\begin{equation}
\tau \propto T,
\label{eq:2-7x}
\end{equation}
and the temperature $T$ is included through the thermal noise $\tau$.

\section{Numerical Results and Discussions}
\label{sec:4}

Using the cell dynamics method presented in the previous section we have simulated the nucleation and growth in a two-dimensional phase field model. The system size is fixed to $512\times512$.  Throughout this work, we set $D=1/2$ and $\psi_{c}=0.2$.  

\begin{figure}[htbp]
\begin{center}
\subfigure[$\epsilon=0.2$ and $1/\tau=40$]{\includegraphics[width=0.90\linewidth]{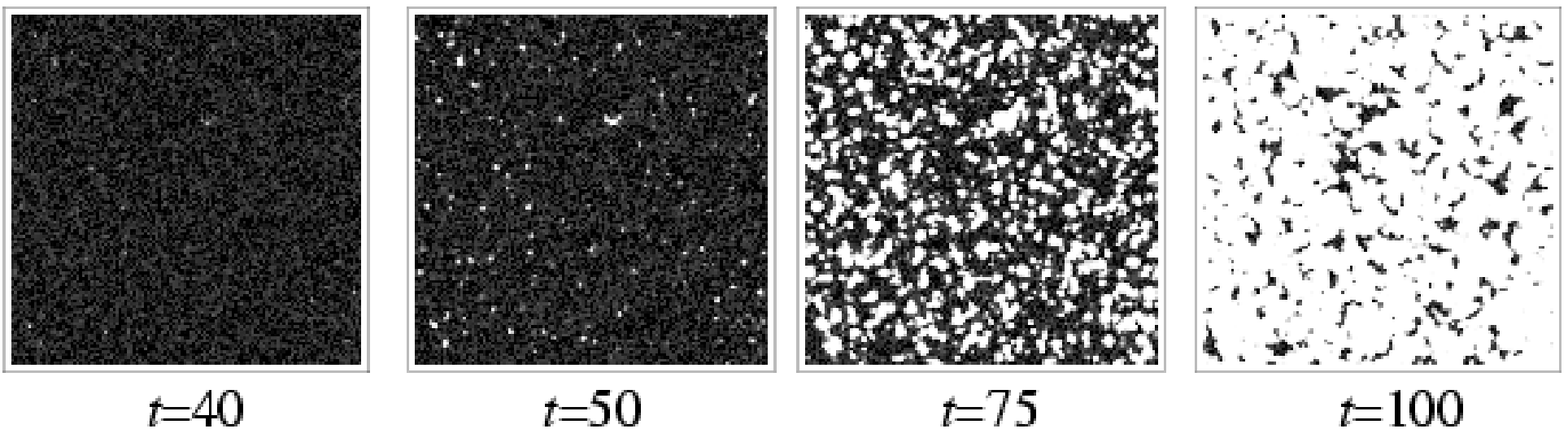}
\label{fig:6a}}
\subfigure[$\epsilon=0.3$ and $1/\tau=40$]{\includegraphics[width=0.90\linewidth]{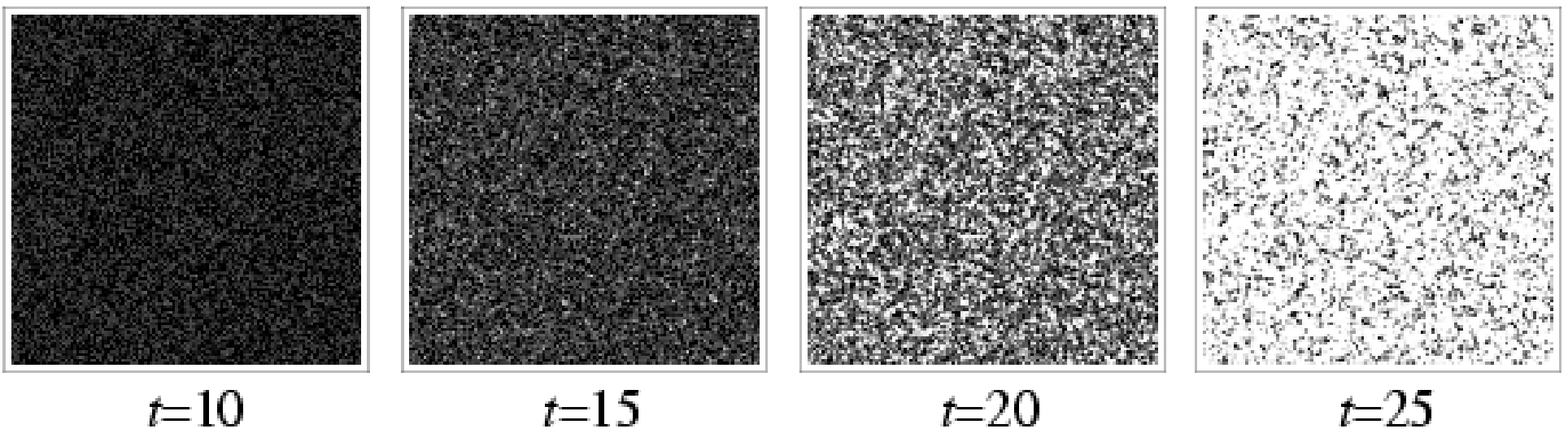}
\label{fig:6b}}
\end{center}
\caption{
A typical evolution pattern of the phase transformation for 512$\times$512 system when (a) the undercooling $\epsilon=0.2$ and (b) $\epsilon=0.3$ at a low temperature $1/\tau=40$. Note that the spinodal is at $\epsilon=1/3$.}
\label{fig:2x}
\end{figure}

Figure~\ref{fig:2x} shows a typical pattern of evolution of a new phase (white) in the metastable old phase (black).  The initial phase is an unstable phase with the order parameter $\psi=0$ (black).  In contrast to the previous method where new nuclear embryos are artificially introduced~\cite{Iwamatsu1,Jou}, near circular nuclei of new phase appear spontaneously from the thermal noise $\tau$ and start to grow (Fig.~\ref{fig:2x}).  As the radius $R_{*}$ of critical nucleus is inversely proportional to the undercooling $\epsilon$ from Eq.~(\ref{eq:3.12x}), the growing nuclei are larger for low undercooling $\epsilon=0.2$ (Fig.~\ref{fig:2x}(a)) than for high undercooling $\epsilon=0.3$ (Fig.~\ref{fig:2x}(b)).  The time scale of phase transformation $t_{\rm inc}$ and $t_{\rm gr}$ are much shorter for high undercooling $\epsilon=0.3$ near the spinodal $\epsilon=1/3$ than for low undercooling $\epsilon=0.2$.

\begin{figure}[htbp]
\begin{center}
\includegraphics[width=0.85\linewidth]{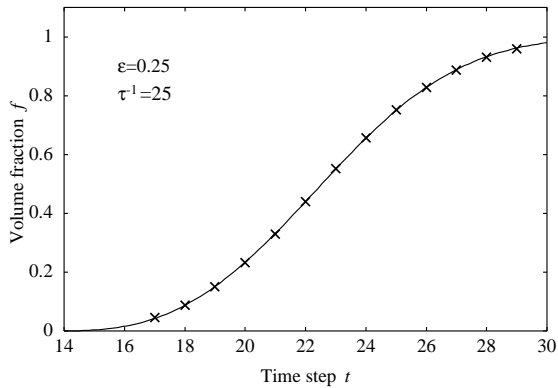}
\end{center}
\caption{
The time evolution of the volume fraction $f$ average over 10 samples for $512\times 512$ system (crosses) compared with the theoretical prediction Eq.~(\ref{eq:3.9}) (solid lines) when $\epsilon=0.25$.}
\label{fig:3x}
\end{figure}

Figure \ref{fig:3x} shows a typical shape of the time evolution of the volume fraction $f$ of transformed volume for two-dimensional $512\times512$ system averaged over 10 samples with different sequences of random number when $\epsilon=0.25$ and $\tau^{-1}=25$.  Since the statistical standard deviations of 10 samples are too small to be visible on the scale of the figure, we did not show the error bars in figure \ref{fig:3x}.  The theoretical curve (solid line) is obtained by the least-square fitting of three parameters, $t_{\rm gr}$, $t_{\rm inc}$ and $m$ in Eq.~(\ref{eq:3.9}) to the simulation data.  Only those simulation data within $0.05\leq f \leq 0.095$~\cite{Mao} is used for least-square fitting to avoid numerical error.  The values for these three parameters differ less than 1\% for larger $1024\times 1024$ system and at most 5\% for smaller $128\times 128$ system. Therefore the use of rather small $512\times512$ system is justified.  In contrast to the previous studies~\cite{Iwamatsu1,Jou}, it can be seen from Fig.~\ref{fig:3x} that the KJMA formula Eq.~(\ref{eq:3.9}) can reproduce the growth curve of $f$ almost perfectly.  

The Avrami exponent $m$, the growth time $t_{\rm gr}$ and the incubation time $t_{\rm inc}$ determined for various temperatures $\tau$ and undercooling $\epsilon$ are summarized in Table \ref{tab:1}.  The Avrami exponents are almost all around the ideal value 3.  Therefore, the classical KJMA picture of nucleation and growth can successfully describe the phase transformation kinetics in our simple phase field model.

\begin{table}[htbp]
\caption{
The Avrami exponent $m$, the growth time $t_{\rm gr}$, and the incubation time $t_{\rm inc}$ determined from direct fitting of Eq.(\ref{eq:3.9}) to simulation data and those determined from the modified Avrami plot Eq.~(\ref{eq:4.1y}).}
\label{tab:1}
\begin{center}
\begin{tabular}{cc|ccc|cc}
\hline
&& \multicolumn{3}{c|}{Direct least-square fit} & \multicolumn{2}{c}{Modified Avrami} \\
\cline{3-7}
\multicolumn{1}{c}{{\raisebox{1.5ex} {\makebox[3em]{$\epsilon$}}}} &
\multicolumn{1}{c|}{\raisebox{1.5ex} {\makebox[3em]{$\tau^{-1}$}}} &
\multicolumn{1}{c}{\makebox[3em]{$m$}} &
\multicolumn{1}{c}{\makebox[3em]{$t_{\rm gr}$}} &
\multicolumn{1}{c|}{\makebox[3em]{$t_{\rm inc}$}} &
\multicolumn{1}{c}{\makebox[3em]{$m$}}&
\multicolumn{1}{c}{\makebox[3em]{$t_{\rm gr}$}} \\
\hline
0.20&  10 &
     	 2.92  &  9.32  &  11.2 & 2.97 & 9.70 \\
    &  15 & 
       2.89  &  12.1  &  15.6 & 2.97 & 12.5 \\
    &  20 &
       2.92  &  15.5  &  19.6 & 2.97 & 15.9 \\
    &  25 & 
       2.97  &  20.1  &  23.6 & 3.04 & 20.7 \\
    &  30 &
       3.12  &  26.8  &  27.3 & 3.15 & 27.3 \\
    & 35  &
       3.24  &  36.2  &  31.3 & 3.30 & 36.8 \\
    & 40 &
       3.43  &  50.5  &  35.5 & 3.44 & 51.0 \\
0.25 &  10 & 
        2.84  &  6.91  &  7.89 & 2.86 & 7.13 \\
    &  15 & 
        2.81  &  7.93  &  10.3 & 2.89 & 6.29 \\
    &  20 & 
        2.78  &  8.94  &  12.2 & 2.82 & 9.17 \\
    &  25 & 
        2.83  &  10.1  &  13.7 & 2.87 & 10.4 \\
    &  30 & 
        2.85  &  11.2  &  15.2 & 2.89 & 11.4 \\
    &  35 &
        2.88  &  12.4  &  16.5 & 2.91 & 12.6 \\
     &  40 & 
        2.91  &  13.6  &  17.8 & 2.98 & 14.0 \\
0.30&  10 &
        2.77  &  5.96  &  6.23 & 2.81 & 6.17 \\
     &  15 & 
        2.71  &  6.53  &  7.90 & 2.74 & 6.72 \\
    &  20 & 
        2.75  &  7.25  &  8.99 & 2.79 & 7.46 \\
    &  25 &
        2.76  &  7.83  &  9.95 & 2.81 & 8.09 \\
     &  30 &  
        2.77  &  8.39  &  10.8 & 2.82 & 8.67 \\
    &  35 &
        2.81  &  8.96  &  11.6 & 2.86 & 9.23 \\
    &  40 & 
        2.83  &  9.50  &  12.3 & 2.88 & 9.76 \\
\hline
\end{tabular}
\end{center}
\end{table}

\begin{figure}[htbp]
\begin{center}
\subfigure[Avrami plot of $f$]{\includegraphics[width=0.90\linewidth]{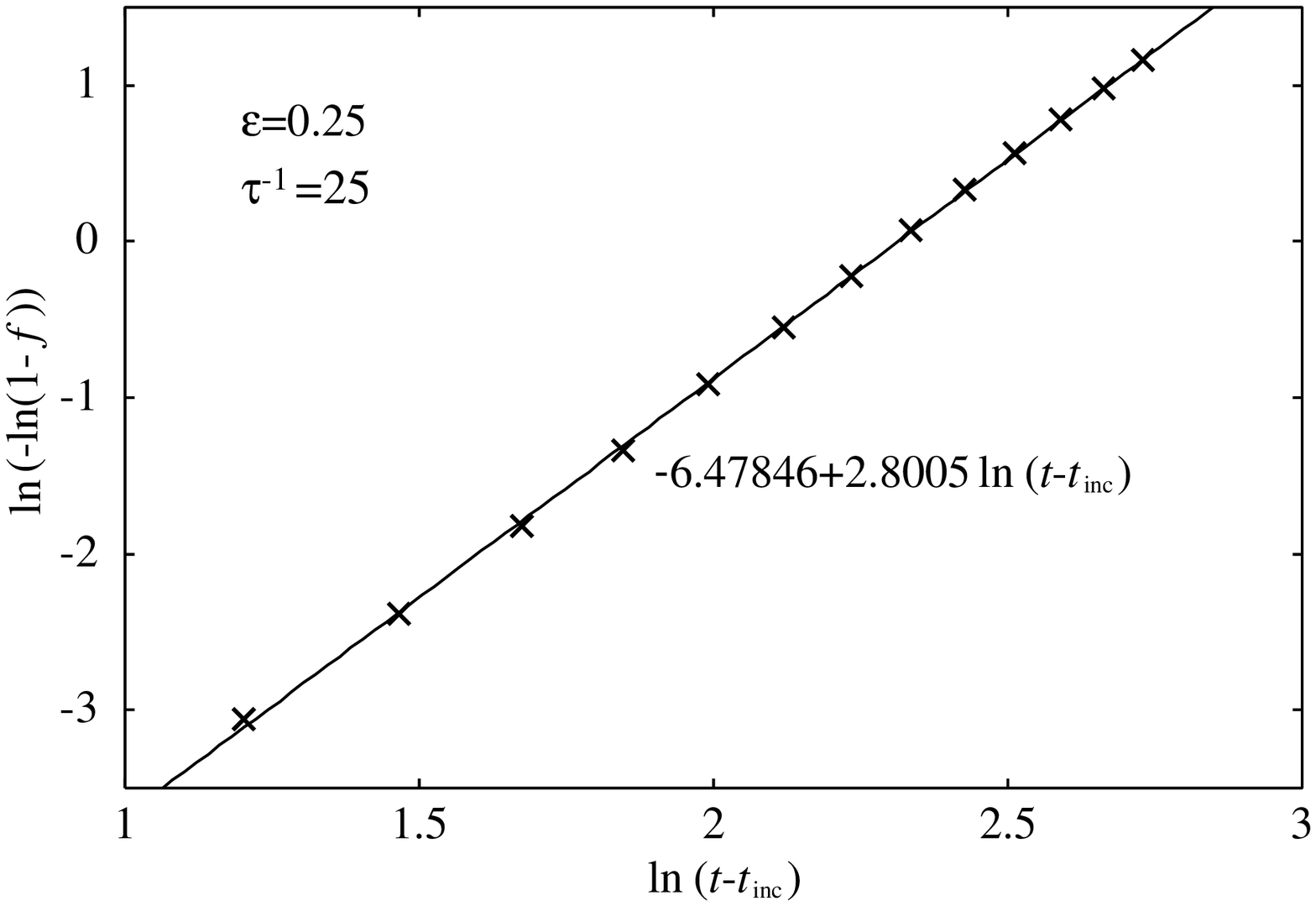}
\label{fig:2a}}
\subfigure[Modified Avrami plot of $f$]{\includegraphics[width=0.90\linewidth]{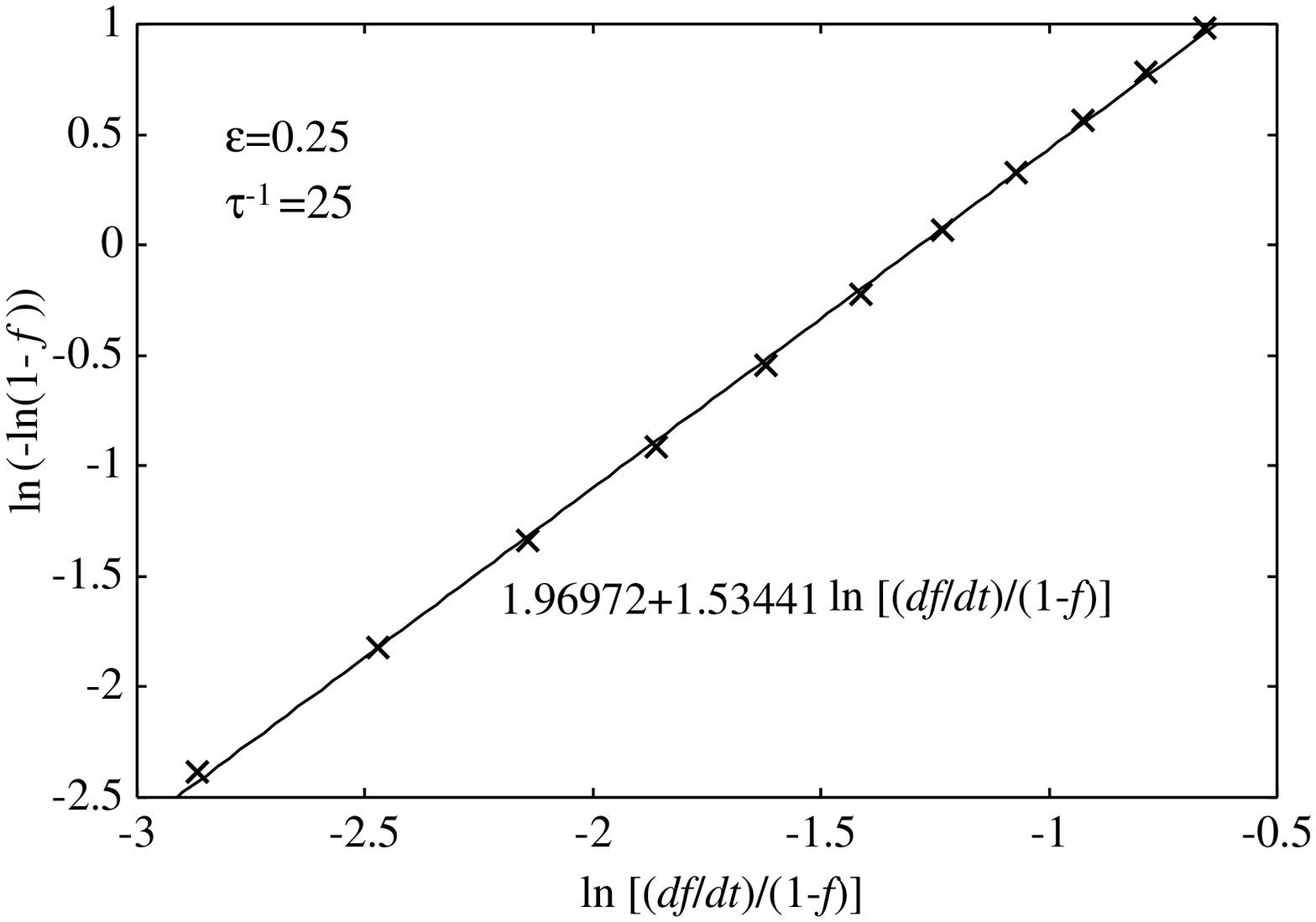}
\label{fig:2b}}
\end{center}
\caption{
(a) The Avrami plot of $f$ (crosses) of Figure \ref{fig:3x}  compared with theoretical predictions Eq.(\ref{eq:3.9}) (solid lines). The incubation time $t_{\rm inc}$ obtained from the least-square fitting in Table~\ref{tab:1} is used to convert to log-plot. (b) The corresponding modified Avrami plot.  The straight lines are obtained by least-square fitting.
}
\label{fig:4x}
\end{figure}

Figure \ref{fig:4x}(a) shows the Avrami plot corresponding to Fig.~\ref{fig:3x}.  From Eq.~(\ref{eq:3.9}) we expect a linear relation
\begin{equation}
\ln \left[-\ln (1-f)\right]=m\ln \left(t-t_{\rm inc} \right) + \mbox{constant}.
\label{eq:4.1}
\end{equation}
between $\ln \left[-\ln (1-f)\right]$ and $\ln \left(t-t_{\rm inc} \right)$ which is the so-called Avrami plot.  The slope gives the Avrami exponent $m=2.8005 \simeq 2.80$ in Fig.~\ref{fig:4x}(a) that is very close to $m=2.83$ for $\epsilon=0.25$ and $\tau^{-1}=25$ in Table \ref{tab:1} deduced directly by fitting the KJMA formula Eq.~(\ref{eq:3.9}). In contrast to the previous approach where the critical nucleus is artificially introduced,~\cite{Iwamatsu1,Jou}, our simulation data in Fig.~\ref{fig:4x}(a) almost perfectly fits to the KJMA formula Eq.~(\ref{eq:4.1}). This is partly due to the fact that the incubation time is unambiguously defined and determined in our simulation, while there is uncertainty of the incubation time in previous works~\cite{Iwamatsu1} as the nucleus of finite size is artificial introduced during the evolution.  The exponents $m$ for other $\epsilon$ and $\tau$ deduced from the slope of the Avrami plot are almost the same as those obtained by direct fitting of Eq.~(\ref{eq:3.9}) to the simulation data tabulated in Table ~\ref{tab:1}.  The results differ at most 10\%. 

It should be noted that the Avrami plot in Fig.~\ref{fig:4x}(a) cannot be used directly to deduce the Avrami exponent $m$ as the plot needs accurate pre-determination of the incubation time $t_{\rm inc}$.  Inappropriate choice of this incubation time is known to lead to non-linear Avrami plot~\cite{Iwamatsu1}. In order to remedy this deficiency, Mao and Altounian~\cite{Mao} proposed modified Avrami plot.  By differentiating Eq.~(\ref{eq:3.9}), we have
\begin{equation}
\frac{df}{dt}=\frac{m}{t_{\rm gr}}\left(\frac{t-t_{\rm inc}}{t_{\rm gr}}\right)^{m-1}\left(1-f\right),
\label{eq:4.1x}
\end{equation}  
which is combined with Eq.~(\ref{eq:3.9}) to give
\begin{equation}
\ln \left[-\ln (1-f)\right]=-\frac{m}{m-1}\ln\left(\frac{m}{t_{\rm gr}}\right)+\frac{m}{m-1}\ln \left[\frac{df/dt}{1-f}\right].
\label{eq:4.1y}
\end{equation}
Eq.~(\ref{eq:4.1y}) is called modified Avrami plot~\cite{Mao}.  Therefore, we expect a linear relation between $\ln \left[-\ln (1-f)\right]$ and $\ln \left[df/dt/(1-f)\right]$. Now, the predetermination of incubation time $t_{\rm inc}$ is unnecessary, and the Avrami exponent can be deduced from the slope $m/(m-1)$, and the growth time $t_{\rm gr}$ can be deduced from the constant $-\left(m/\left(m-1\right)\right)\ln\left(m/t_{\rm gr}\right)$. 

This modified Avrami plot in Figure \ref{fig:4x}(b) clearly shows a predicted linear relation.  Since we use cell dynamics method, we have replaced the differentiation $df/dt$ by the finite difference in Eq. (\ref{eq:4.1y}).  The Avrami exponent deduced from this modified Avrami plot $m=1.53441/(1.53441-1)\simeq 2.87$ is very close to $m=2.83$ for $\epsilon=0.25$ and $\tau^{-1}=25$ in Table \ref{tab:1} deduced directly by fitting KJMA formula Eq. (\ref{eq:3.9}) to simulation data.  Therefore, the validity of KJMA picture in our phase field model is further confirmed. The Avrami exponents $m$ and the growth times $t_{\rm gr}$ deduced from this modified Avrami plot differs at most a few percent from those deduced from direct fitting (Table~\ref{tab:1}).

Although, the validity of KJMA picture represented by the KJMA formula in Eq.~(\ref{eq:3.9}) in our phase field model is confirmed, the Avrami exponent $m$ obtained by fitting to the simulation data shows slight deviation from the ideal value $m=3$ as shown in Table~\ref{tab:1}, which means that the time dependence of nucleation rate $I\simeq I_{0}t^{\alpha}$ and interfacial velocity $v\simeq v_{0}t^{\beta}$ with $\alpha\neq 0$ or $\beta\neq 0$ exist.  

\begin{figure}[htbp]
\begin{center}
\includegraphics[width=0.90\linewidth]{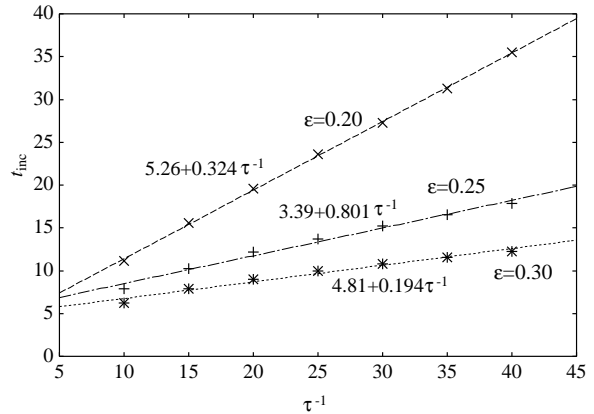}
\end{center}
\caption{
The incubation time $t_{\rm inc}$ as the function of the inverse temperature $1/\tau\propto 1/T$.  The incubation times do not increase exponentially but they increase linearly, and are roughly proportional to $1/\tau$ as predicted from Shneidman-Weinberg theory~\cite{Shneidman1}in Eq.~(\ref{eq:3.11x}). The straight lines are obtained by least-square fitting.
  }
\label{fig:5x}
\end{figure}

Figure \ref{fig:5x} shows the incubation time $t_{\rm inc}$ estimated from the least-square fitting of KJMA curve in Eq.~(\ref{eq:3.9}) to the simulation data and tabulated in Table \ref{tab:1} as the function of temperature parameter $1/\tau$ for various undercooling $\epsilon$.  In contrast to the previous report for the Ising system~\cite{Shneidman3}, a fortunate cancellation of the first and the second term of Eq.~(\ref{eq:3.7}) does not occur and we have rather long incubation times in our phase-field model (Table~\ref{tab:1}).

Usually the exponential Arrhenius-type temperature dependence~\cite{Nagpal,Kalb} 
\begin{equation}
t_{\rm inc} \propto \exp\left(\frac{W_{*}}{k_{\rm B}T}\right)
\label{eq:4.2x}
\end{equation} 
is assumed to analyze experimental incubation time.  However, the incubation times obtained in Table~\ref{tab:1} do not increase exponentially but they increase linearly in Fig.~\ref{fig:5x} . The curve shows a nearly linear behavior expected from Eq.(\ref{eq:3.11x}):
\begin{equation}
t_{\rm inc}\propto \frac{W_{*}}{m k_{\rm B}T} \propto \frac{1}{\tau}
\label{eq:4.2}
\end{equation}
as the function of the inverse temperature $1/\tau\propto 1/T$ with a slope that is inversely related to the undercooling $\epsilon$ as the energy barrier $W_{*}$ is inversely proportional to $\epsilon$ from Eq.~(\ref{eq:3.12y})~\cite{Shneidman1}.  Therefore when the undercooling is low ($\epsilon=0.2$), the slope is steep and the incubation time increases rapidly as the temperature is lowered.  Our numerical results in Fig.~\ref{fig:5x} seems to suggest that the theory of Shneidman and Weinberg theory~\cite{Shneidman1} is qualitatively correct in our phase field model.

\begin{figure}[htbp]
\begin{center}
\includegraphics[width=0.90\linewidth]{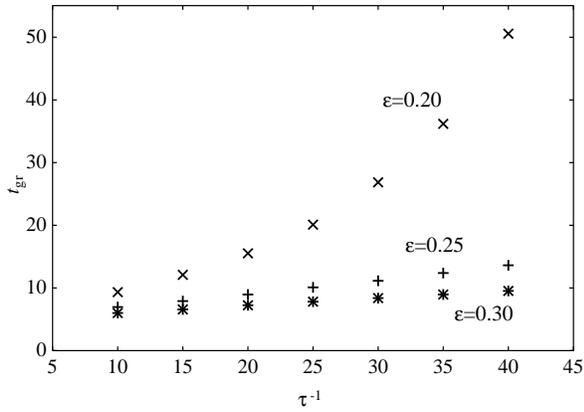}
\end{center}
\caption{
The growth time $t_{\rm gr}$ as the function of the inverse temperature $1/\tau\propto 1/T$. }
\label{fig:6x}
\end{figure}

Figure~\ref{fig:6x} shows the temperature dependence of the growth time $t_{\rm gr}$.  As intuitively expected, the growth time is an increasing function of the inverse temperature $\tau^{-1}$.  It increases as the temperature is lowered because nucleation rate becomes low.  We should note that in our phase-field model, the interfacial velocity $v_{0}\simeq v$ does not depend on the temperature from Eq.~(\ref{eq:3.13}).  Therefore, the temperature dependence of $t_{\rm gr}$ in Eq.~(\ref{eq:3.10}) comes mainly from the temperature dependence of the nucleation rate $I_{0}\simeq I_{s}$.

\begin{figure}[htbp]
\begin{center}
\includegraphics[width=0.90\linewidth]{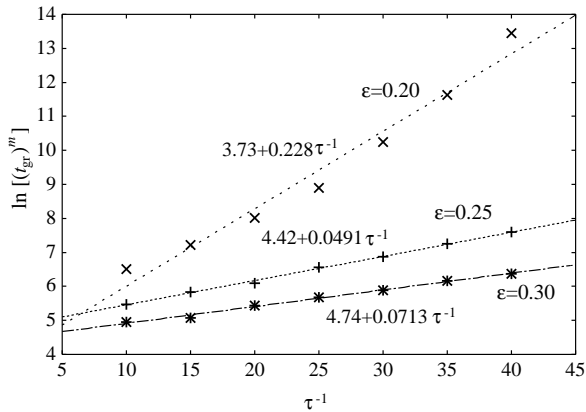}
\end{center}
\caption{
The logarithm of the growth time $\ln\left[\left(t_{\rm gr}\right)^{m}\right]$ proportional to that of the nucleation rate $\ln\left[I_{s}\right]$ as the function of the inverse temperature $1/\tau\propto 1/T$.  The logarithm of the nucleation rate $\ln\left[I_{s}\right]$ are roughly proportional to $1/\tau$ in accordance with the CNT. The straight lines are obtained by least-square fitting.  }
\label{fig:7x}
\end{figure}

In fact, from Eq.~(\ref{eq:3.10}) we expect an activation-type temperature dependence 
\begin{equation}
\left(t_{\rm gr}\right)^{m} \propto I_{s}^{-1} \propto \exp\left(\frac{W_{*}}{k_{B}T}\right)
\label{eq:4.3}
\end{equation}
if the classical nucleation theory (CNT) is valid.  Therefore, the Arrhenius plot of the growth time $\left(\tau_{\rm gr}\right)^{m}$ versus inverse temperature $\tau^{-1}$ should follow a straight line.  Figure \ref{fig:7x} shows the Arrhenius (semi-log) plot of the growth time $\left(\tau_{\rm gr}\right)^{m}$.  Both $\tau_{\rm gr}$ and $m$ are those obtained from the least-square fitting of Eq.~(\ref{eq:3.9}) and are tabulated in the first two columns of Table \ref{tab:1}.  The curves are almost straight lines expected from CNT in Eq.~(\ref{eq:4.3}) for high undercoolings $\epsilon=0.25$ and $\epsilon=0.30$. The classical picture of activation-type nucleation rate of CNT seems valid for high undercooling in our phase-field model.  Then the total time scale of the phase transformation $t_{\rm inc}+t_{\rm gr}$ that is the sum of the incubation time $t_{\rm inc}$ and the growth time $t_{\rm gr}$, will not be described by a simple formula.   Neither Eq.~(\ref{eq:4.3}) predicted from CNT nor Eq.~(\ref{eq:4.2}) of  Shneidman and Weinberg theory~\cite{Shneidman1} alone cannot describe the temperature dependence of total transformation time.

The curve deviated, however, from a straight line for the low undercooling $\epsilon=0.2$ in Fig.~\ref{fig:7x}. Therefore, our numerical results seem to indicate that the CNT in Eq.~(\ref{eq:3.11}) cannot be used for the phase-field model when the undercooling is low. Looking at the curve in Fig.~\ref{fig:7x} for $\epsilon=0.2$ closely, we note that the curve start to deviate from a straight line when the inverse temperature becomes large $\tau^{-1}>30$ when the Avrami exponent $m$ also becomes larger than the ideal value 3 (Table \ref{tab:1}).  In fact, our cell dynamics code can not simulate the nucleation process correctly for low undercooling.  It should be noted that we have artificially introduced an ad hock cutoff $\psi_{c}=0.2$ to prevent the thermal fluctuation $\xi(t,n)$ and the extra nucleation in our simulation.  This choice $\psi_{c}=0.2$ may not be appropriate since the critical nucleus is large and the energy barrier is high for low undercooling $\epsilon=0.2$.  In fact, the radius of critical nucleus become $R_{*}=\sigma/\Delta h\simeq 0.8$ from Eqs.~(\ref{eq:3.12x}) and (\ref{eq:2-9}) as $\sigma = 1/24\simeq 0.04$~\cite{Iwamatsu4} and $\Delta h = 0.2/4=0.05$ when $\epsilon=0.2$, and roughly $\pi R_{*}^{2}\simeq 2$ to $3$ neighboring cells should be transformed simultaneously in order to form a critical nucleus that can continue to grow.  Then the small change of cutoff $\psi_{c}$ greatly influence the nucleation process and may affect the phase transformation dynamics.  Therefore the deviations of simulation data from the prediction of CNT in Fig.~\ref{fig:7x} when the undercooling $\epsilon=0.2$ could be due to the inappropriate choice of cutoff parameter $\psi_{c}$ and/or the limitation of controlling nucleation event by a single cutoff parameter.  

\begin{figure}[htbp]
\begin{center}
\includegraphics[width=0.90\linewidth]{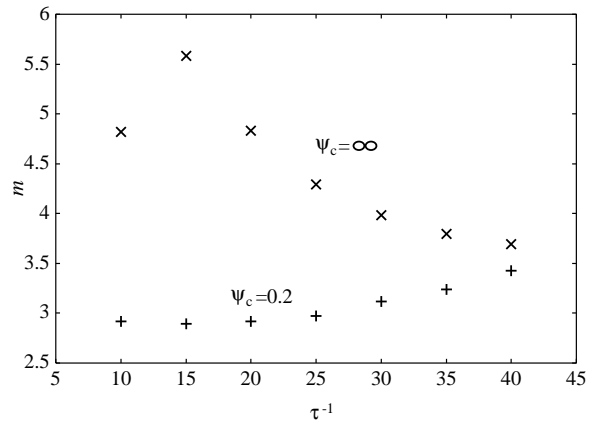}
\end{center}
\caption{
The Avrami exponent $m$ as the function of the inverse temperature $1/\tau\propto 1/T$.  The Avrami exponent $m$ is much larger when the cutoff $\psi_{c}$ is not introduced ($\psi_{c}=\infty$) than when $\psi_{c}=0.20$.}
\label{fig:8x}
\end{figure}

When the cutoff parameter is not introduced ($\psi_{c}=\infty$), the thermal fluctuation $\xi(t,n)$ and the extra nucleation may occurs even in the region where the nucleation has already started.  Then the nucleation rate should be enhanced.  The nucleation rate increases as more and more materials are transformed.  The nucleation rate becomes the increasing function of time $I(t)\simeq I_{0}t^{\alpha}$ with non-zero exponent $\alpha\neq 0$, which will lead to a larger Avrami exponent $m$ from Eq.~(\ref{eq:3.9a}).  Actually the Avrami exponent $m$ in Table \ref{tab:1} becomes larger than the ideal value $3$ for low temperatures $\tau^{-1}\geq 30$ at the low undercooling $\epsilon=0.2$.  Therefore the artificial cutoff $\psi_{c}$ alone may not control the nucleation event in our phase field model properly.  

Figure \ref{fig:8x} compares the Avrami exponents $m$ calculated when $\psi_{c}=0.2$ with those calculated when $\psi_{c}=\infty$ (without cutoff).  The Avrami exponent $m$ becomes much larger than ideal value $m=3$ when $\psi_{c}=\infty$ as expected. However, since we can still deduce not only the Avrami exponent $m$ but the incubation time $t_{\rm inc}$ and the growth time $t_{\rm gr}$, the KJMA picture of phase transformation represented by Eq.~(\ref{eq:3.9}) is still qualitatively valid even when $\psi_{c}=\infty$.  Therefore, qualitative picture of nucleation and growth will be still valid even if we do not introduce cutoff $\psi_{c}$.

So far, such a thermal fluctuation is expected to initiate nucleation of patter formation in phase field model~\cite{Valls,Castro} or cell dynamics~\cite{Oono,Puri,Ren}, and is uncritically used without any special care such as our cutoff in this paper.  Our phase field simulation for the most basic process of nucleation and growth clearly showed that such a naive expectation is only qualitatively valid and may not lead to quantitatively correct description of the dynamics of phase transformation.

\section{Conclusion}
\label{sec:sec4}

In this report, we used the cell dynamics method~\cite{Iwamatsu1} to study the whole dynamics of phase transformation from the homogeneous nucleation to growth in a simple phase field model.  We found that the Kolmogorov-Johnson-Mehl-Avrami (KJMA) scenario of nucleation and growth is correct in this phase-field model.  Specifically, the evolution of the volume fraction of transformed volume follows the KJMA formula.  By fitting the KJMA formula to the simulation data, we could deduce not only the Avrami exponent $m$ but the incubation time $t_{\rm inc}$ and the growth time $t_{\rm gr}$. So far as the present author knows, there has been virtually no detailed quantitative study of the incubation time.  Therefore our report is probably the first quantitative report of the theoretical study of the incubation time $t_{\rm inc}$.

The Avrami exponent $m$ is found to be very close to the ideal value $m=3$. The incubation time $t_{\rm inc}$ is inversely proportional to the absolute temperature and is qualitatively in accord with the theory of Shneidman-Weinberg~\cite{Shneidman1}.  This conclusion would be valid for real materials unless some activation-type temperature dependence comes in through, for example, diffusion-limited growth.  In fact, many experimental workers~\cite{Nagpal,Kalb} successfully analyzed their experimental data assuming exponential activation-type temperature dependence of incubation time without paying much attention to the Shneidman-Weinberg theory.  However, this could partly due to the ambiguity of the definition of experimentally determined incubation time that could be actually the growth time. In our study we have separated the incubation time and growth time unambiguously using the KJMA formula for the volume fraction.  It turns out that both the incubation time and the growth time are the same order of magnitude in the phase-field model.  Therefore, similar clear discrimination between the incubation time and the growth time, if possible, would be necessary to analyze experimental data quantitatively.

The temperature dependence of nucleation rate $I$ deduced from the growth time $t_{\rm gr}$ follows the classical nucleation theory except at low undercooling. The error at the low undercooling is caused from the difficulty of controlling nucleation event in phase field model.  An ad hoc cutoff $\psi_{c}$ to prevent thermal fluctuation and extra nucleation in the region where the phase transformation has already started cannot work well for the low undercooling.  Therefore, a blind use of thermal fluctuation in phase field model and/or cell dynamics method needs caution.  It may not give a quantitatively correct dynamics of phase transformation.  Our numerical simulation has unambiguously shown that an inappropriate choice of this cutoff parameter will lead to quantitatively incorrect Avrami exponent $m$.  

In conclusion, we have shown that the most basic dynamics of phase transformation in the phase field model follows the scenario of Kolmogorov-Johanson-Mehl-Avrami.  Furthermore, we have demonstrated that the incubation time $t_{\rm inc}$ and the growth time $t_{\rm gr}$ show different temperature dependence.  Therefore, the total phase transformation time $t_{\rm inc}+t_{\rm gr}$ will not be described by a simple single formula.  If, however, $t_{\rm gr}$ is dominant, {that is expected when the temperature is lowered,} the temperature dependence of phase transformation time follows usual activation-type formula of CNT.  On the other hand, if $t_{\rm inc}$ is dominant, the phase transformation time is not activation-type but is inversely proportional to the absolute temperature~\cite{Shneidman1}.

Finally, in contrast to the standard cellular automaton~\cite{Marx,Hesselbarth} and the lattice model ~\cite{Rollett,Castro2} where artificial evolution processes are introduced algorithmically, the evolution in the phase-field model is driven by the free energy and the thermal noise, and is completely free from artificial parameters.  Therefore, the phase-field model with cell dynamics method is powerful method to study the various phase transformation scenarios, in particular, of the time evolution including the incubation time and the growth time without introducing many unknown and uncontrollable parameters.  For example, various scenarios of heterogeneous nucleation are already clarified~\cite{Iwamatsu4} by this phase-field model with cell dynamics method.

\begin{acknowledgments}
The author is grateful to Dr. M. Nakamura for his helpful comments and his help during initial stage of this work.  He is also grateful to the reviewer for his/her careful examination of the manuscript and useful comments.

\end{acknowledgments}

\end{document}